\documentstyle[prl,aps,twocolumn,floats,amssymb,epsfig]{revtex}

\def\bbN{{\mathbb N}}

\def\bbZ{{\mathbb Z}}
\def\su{{\rm{su}}}
\def\u{{\rm{u}}}
\def\SU{{\rm{SU}}}
\def\U{{\rm{U}}}
\newcommand{\mb}[1]{\ifmmode#1\else\mbox{$#1$}\fi}

\newcommand\al{\mb{\alpha}}

\newcommand\be{\mb{\beta}}
\newcommand\ga{\mb{\gamma}}
\newcommand\de{\mb{\delta}}

\newcommand\Om{\mb{\Omega}}

\newcommand\calH{\mb{{\cal H}}}

\newcommand\calM{\mb{{\cal M}}}

\renewcommand{\ll}[1]{\label{#1}}

\newcommand{\beq}{\begin{equation}}
\newcommand{\eeq}{\end{equation}}
\newcommand{\nn}{\nonumber}
\newcommand{\bea}{\begin{eqnarray}}
\newcommand{\eea}{\end{eqnarray}}

\newcommand{\mod}[1]{\mid \!\! {#1} \!\! \mid}

\newcommand{\tr}{\mb{{\rm tr}\,}}

\newcommand{\x}{\mb{\times}}
\newcommand{\rhat}{\mb{\hat{\bm r}}}

\newcommand{\Ad}{{\rm Ad}}
\newcommand{\ad}{{\rm ad}}
\newcommand{\diag}{{\rm diag}}
\newcommand{\gsim}
{\raise.3ex\hbox{$\;>$\kern-.75em\lower1ex\hbox{$\sim$}$\:$}}
\newcommand{\lsim}
{\raise.3ex\hbox{$\;<$\kern-.75em\lower1ex\hbox{$\sim$}$\:$}}
\newcommand{\ts}{\textstyle}

\newcommand{\half}{\mb{\ts \frac{1}{2}}}

\newcommand{\third}{\mb{\ts \frac{1}{3}}}
\newcommand{\twothird}{\mb{\ts \frac{2}{3}}}

\newcommand{\bm}[1]{{\mbox{\boldmath $#1$}}}

\newcommand{\rmC}{{\rm C}}

\newcommand{\rmI}{{\rm I}}
\newcommand{\rmY}{{\rm Y}}
\newcommand{\m}{{\mbox{$$-$$}}}

\begin{document}
\draft


\twocolumn[\hsize\textwidth\columnwidth\hsize\csname @twocolumnfalse\endcsname
\title{Some simpler analogues of the dual standard model and their relation\\
to Bais' generalization of the Montenon-Olive conjecture} 
\author{Nathan\  F.\  Lepora\footnotemark}
\date{Submitted 23 October 2001; accepted 25 November 2001}
\maketitle

\begin{abstract}
We show that the correspondence between SU(5) monopoles and 
the elementary particles,
which underlies the construction of a dual standard model,
has some simpler analogues associated with the strong, weak and hypercharge
interactions. We then discuss how these analogues 
relate to Bais' generalization of
the Montenon-Olive conjecture and find the representations of
the monopoles under the dual gauge group; these representations
agree with those of the elementary particles.
\end{abstract}
\pacs{Published in: Phys. Lett. {\bf B}524 (2002) 388.}]


Recently, Vachaspati has discovered a remarkable correspondence between the 
elementary particles and the monopoles
from Georgi-Glashow SU(5) unification~\cite{vach95}. He found that 
the magnetic charges of the five stable monopoles from 
\beq
\ll{unif} 
\SU(5) \rightarrow \SU(3)_\rmC\x\SU(2)_\rmI\x\U(1)_\rmY/\bbZ_6 
\eeq 
are identical to the electric charges of the five particle multiplets
in one standard model generation (see table~\ref{tab1}). 
This finding 
suggests the elementary particles may have a description as monopoles 
in a dual SU(5) theory, somewhat in analogy with the Skyrme model of nucleons.

For such an approach to be successful,
other aspects of the standard model should naturally arise as properties 
of the SU(5) classical solutions. At present the investigation of this is
an ongoing task, although 
many features are falling into place. For instance, both spin and confinement 
can occur~\cite{vach95} and the dual gauge couplings successfully 
unify~\cite{meunif}. 

In this letter, we discuss some simpler analogues of the
monopole-elementary particle correspondence that are both interesting in 
themselves and can aid study of the dual standard model.
These correspondences occur in the following symmetry breakings
\bea
\ll{su4}\SU(4)&\rightarrow& \SU(3)_\rmC\x\U(1)_\rmY/\bbZ_3,\\
\ll{su3}\SU(3)&\rightarrow& \SU(2)_\rmI\x\U(1)_\rmY/\bbZ_2,\\
\ll{su2}\SU(2)&\rightarrow& \U(1)_\rmY
\eea
and relate to the subset of elementary particles
charged only under the appropriate residual gauge symmetry. 

Later we use these models to relate the monopole-elementary particle 
correspondences to Bais' generalization~\cite{bais} of the 
Montenon-Olive conjecture~\cite{mont},
which is itself closely related to the Goddard, Nuyts and Olive 
conjecture~\cite{gno}. These conjectures represent a well known and accepted 
description of the magnetic sector of non-Abelian gauge theories.
In consequence, we can check that the representations (and not just the 
charges) of the monopoles and elementary particles match. 

To see how these models relate to the full SU(5) theory, we recall some 
details of the SU(5) monopole spectrum arising from (\ref{unif}). These
\pagebreak
monopoles are specified by their asymptotic magnetic field 
\beq
\label{5}
\bm B \sim \frac{1}{2e}\frac{\bm\rhat}{r^2}M(\Om),\hspace{2em}
M(\hat\bm z)=M
\eeq
with $M(\Om)$ covariantly constant. Then Gardner and Harvey showed that
the stable monopoles have a spectrum
for scalar masses much greater than gauge masses~\cite{gard84}:

\begin{table}[h]
\caption{$\SU(5)\rightarrow\SU(3)\x\SU(2)\x\U(1)/\bbZ_6$ Monopoles.}
\begin{tabular}{cccccc}
topology $n$ & $m_{\rm C}$ & $m_{\rm I}$ & $m_{\rm Y}$ & diag $M$ & 
multiplet \\
\hline
1 & 1 & $\half$ & $\third$ & $(0,0,1,\m 1,0)$ & $(u, d)_L$  \\
2 & $\m$ 1 & 0 & $\twothird$ & $(0,1,1,\m 1,\m 1)$ & $\bar{d}_L$ \\
3 & 0 & $\m\half$ & 1 & $(1,1,1,\m 2,\m 1)$ & $(\bar{\nu}, \bar{e})_R$ \\
4 & 1 & 0 & $\ts\frac{4}{3}$ & $(1,1,2,\m 2,\m 2)$ & $u_R$ \\
5 & - & - & - & - & - \\
6 & 0 & 0 & 2 & $(2,2,2,\m 3,\m 3)$ & $\bar{e}_L$ \\
$\cdots$ & - & - & - & - & - \\
\end{tabular}
\label{tab1}
\end{table}

\noindent
Here the individual magnetic charges are defined by 
$M = m_\rmC T_\rmC + m_\rmI T_\rmI + m_\rmY T_\rmY$ with a convenient
choice of generators being
\bea
T_{\rm C}={\rm diag}(\m\third, \m\third, \twothird, 0,0),\hspace{1em}
T_{\rm I}={\rm diag}(0,0,0, 1, \m 1),\nn\\
T_{\rm Y}={\rm diag}(1,1,1,\m\ts\frac{3}{2},\m\ts\frac{3}{2}).\hspace{5em}
\eea
Also included in table \ref{tab1} are the standard model multiplets
with identical charges to the monopoles.

Notice the monopoles with $n=5$ and $n\geq 7$ are unstable.
This is because the monopole-monopole potential $V(r)$ monotonically
tends to zero at infinity, while~\cite{vach95,gard84}
\bea
\label{7}
V(0)=\frac{1}{4\al}\left[ \tr(T_\rmC^2) m^{}_\rmC m_\rmC' \mu^{}_\rmC +
\tr(T_\rmI^2) m^{}_\rmI m_\rmI' \mu^{}_\rmI\right.\hspace{2em}\nn \\
\left.+\,\tr(T_\rmY^2) m^{}_\rmY m_\rmY' \mu^{}_\rmY \right],
\eea
where the $\mu$'s relate to the scalar boson masses and
$\al=e^2/4\pi$. Only the indicated
monopoles in table~\ref{tab1} have all possible fragmentations satisfying 
$V(0)<0$ for suitable values of $\mu$, implying only these monopoles are 
stable.

In addition to the monopoles in table~\ref{tab1}, there are anti-monopoles
with magnetic charges
$(\m m_\rmC,\m m_\rmI,\m m_\rmY)$ --- these correspond to
$(\bar u,\bar d)_R$, $d_R$, $(\nu,e)_L$, $\bar u_L$ and $e_R$.

The monopoles also have some degeneracy for their diagonal
generators: colour has a three-fold degeneracy
$T_\rmC^r=\diag(\m\third,\m\third,\twothird,0,0)$, 
$T_\rmC^g=\diag(\m\third,\twothird,\m\third,0,0)$ and
$T_\rmC^b=\diag(\twothird,\m\third,\m\third,0,0)$;
while weak isospin has a two-fold degeneracy $T^\pm_\rmI=\pm T_\rmI$. 

We now find the monopole spectra for the symmetry breakings~(\ref{su4}) 
to (\ref{su2}) by applying the argument between (\ref{5})
and (\ref{7}) to these models. Because they have residual symmetries 
contained within that of (\ref{unif}), similar patterns of monopoles 
to table~\ref{tab1} are expected. Specifically:

\noindent (i) $\SU(4)\rightarrow\SU(3)_\rmC\x\U(1)_\rmY/\bbZ_3$:
For this we write 
\bea
M=m_\rmC T_\rmC + m_\rmY T_\rmY,\hspace{5em}\\
T_\rmC = \diag(\m\third,\m\third,\twothird,0),
\hspace{1em}T_\rmY=\diag(\half,\half,\half,\m\ts\frac{3}{2}).
\eea
Then $V(0)=(m_\rmC m_\rmC'\mu_\rmC + m_\rmY m_\rmY'\mu_\rmY)/4\al$ implies 
the stable monopoles have a correspondence for suitably small $\mu_\rmY$:

\begin{table}[ht]
\caption{$\SU(4)\rightarrow\SU(3)_\rmC\x\U(1)_\rmY/\bbZ_2$ monopoles.}
\begin{tabular}{ccccc}
topology $n$ & 
$m_{\rm C}$ & $m_{\rm Y}$ & diag $M$ & multiplet \\
\hline
1 & 1 & $\twothird$ & $(0,0,1,\m 1)$ & $\bar{d}_L$ \\
2 & $\m 1$ & $\ts\frac{4}{3}$ & $(1,1,0,\m 2)$ & $u_R$ \\
3 & 0 & 2 & $(1,1,1,\m 3)$ & $\bar{e}_L$ \\
$\cdots$ & - & - & - & - \\
\end{tabular}
\label{tab4}
\end{table}
\noindent
In addition, there are anti-monopoles, which correspond to 
$d_R$, $\bar u_L$ and $e_R$. There is also a three-fold colour
degeneracy of the $n=1,2$ monopoles.

Therefore, by isolating just the colour and hypercharge gauge symmetries,
a correspondence with the standard model weak isospin singlets is obtained.

\noindent (ii) $\SU(3)\rightarrow\SU(2)_\rmI\x\U(1)_\rmY/\bbZ_2$:
Now we write
\bea
M=m_\rmI T_\rmI + m_\rmY T_\rmY,\hspace{4em}\\
T_\rmI = \diag(1,\m 1,0),\hspace{1em}T_\rmY=\diag(\half,\half,\m 1).
\eea
Then $V(0)=(m_\rmI m_\rmI'\mu_\rmI +m_\rmY m_\rmY'\mu_\rmY)/4\al$
implies the stable monopoles have a 
correspondence for suitably small $\mu_\rmY$:

\begin{table}[h]
\caption{$\SU(3)\rightarrow\SU(2)_\rmI\x\U(1)_\rmY/\bbZ_2$ monopoles.}
\begin{tabular}{ccccc}
topology $n$ & $m_{\rm I}$ & $m_{\rm Y}$ & diag $M$ & multiplet \\
\hline
1 & $\m\half$ & 1 & $(0,1,\m 1)$ & $(\bar{\nu}, \bar{e})_R$ \\
2 & 0 & 2 & $(1,1,\m 2)$ & $\bar{e}_L$ \\
$\cdots$ & - & - & - & - \\
\end{tabular}
\label{tab3}
\end{table}
\noindent
In addition, there are anti-monopoles, which correspond to 
$(\nu, e)_L$ and $e_R$. 
There is also a two-fold gauge degeneracy of the $n=1$ monopoles 
upon taking $T_\rmI^\pm=\pm T_\rmI$.

Therefore isolating the electroweak gauge symmetry gives a 
correspondence with the standard model colour singlets (leptons).

\noindent (iii) $\SU(2)\rightarrow\U(1)_\rmY$: Finally, we write
\beq
M=m_\rmY T_\rmY,\hspace{2em}
T_\rmY=\diag(\half,\m \half)
\eeq
Then $V(0)=m_\rmY m_\rmY'\mu_\rmY/4\al$ implies only the
't~Hooft-Polyakov monopole is stable, with a correspondence:\\
\vspace{-1em}

\begin{table}[h]
\caption{$\SU(2)\rightarrow\U(1)_\rmY$ monopoles.}
\begin{tabular}{cccc}
topology $n$ & $m_{\rm Y}$ & diag $M$ & multiplet \\
\hline
1 & 2 & $(1,\m 1)$ & $\bar{e}_L$ \\
$\cdots$ & - & - & - \\
\end{tabular}
\label{tab2}
\end{table}
\noindent
In addition, there is an anti-monopole with $m_\rmY=-2$, which has the 
same charges as $e_R$.
Thus isolating the hypercharge gauge symmetry gives a
correspondence with the only weak isospin and colour singlet.

Let us now discuss how the above relates to non-Abelian 
electric-magnetic duality. For this discussion,
it will be necessary to recall some results about general magnetic monopoles. 
In particular, we consider a simple compact gauge
group $G$ with a scalar field $\phi$ in the adjoint representation so
the residual gauge symmetry satisfies $\Ad(H)\phi_0=\phi_0$ 
with $\phi_0$ the vacuum expectation value.

The root sets $\Phi$ of $G$ and $H$ will be important, 
so we consider a basis $\{T_1,\cdots,T_r\}$ for a maximal Abelian subalgebra
of $H$. Then the roots $\bm\al$ satisfy
\beq
i\,\ad(\bm T)E_\al = \bm\al E_\al,
\eeq
Without loss of generality $\phi_0 = \bm f \cdot \bm T$,
which implies the gauge bosons have a spectrum and masses~\cite{bais}
\bea
\label{mgauge}
W_\mu=\sum_{\al\in\Phi(G)} W^{\al}E_{\al},\hspace{2em}
(M_W)_{\pm\al} = e \mod{\bm f \cdot \bm\al}
\eea
with $W^{-\al}=(W^{\al})^\dagger$ and normalization
$[E_\al,E_{-\al}]=\bm\al\cdot\bm T$. 

The magnetic monopoles have the asymptotic form
\beq
\bm B \sim \frac{1}{2e}\frac{\bm\rhat}{r^2}M(\Om),\hspace{2em}
M(\hat\bm z)=M.
\eeq
Without loss of generality $M=\bm m\cdot\bm T$ --- then $M$ satisfies
a topological quantization with solutions~\cite{gno,engle}
\beq
\exp(2\pi iM)=1 
\hspace{1em}\Rightarrow\hspace{1em}
\bm m = \sum_a n_a \bm\be_{(a)}^*,\hspace{.6em}
n_a\in\bbN,
\eeq
where $\bm\be^*_{(a)}=\bm\be_{(a)}/\be_{(a)}^2$ are the duals of the
simple roots $\bm\be_{(a)}$ of $H$, 
which span $\Phi(H)$ with integer coefficients.

An important observation by Goddard, Nuyts and Olive~\cite{gno}
is there is a dual group $H^*$ defined by the roots
\beq
\label{H*}
\Phi(H^*) = 
\left\{\bm\be^*/N:\bm\al\in\Phi(H)\right\}
\eeq
for some constant $N$ depending upon the group. Since particle
multiplets have charges that are weights of an associated symmetry group,
one concludes the magnetic monopoles form multiplets of $H^*$. 
By analogy with electric interactions, this dual group $H^*$ 
should then describe the magnetic gauge interactions.

Bais, following Montenon and Olive,
gave further support for this conjecture by considering the theory
in the BPS limit. In particular, he considered those monopoles with 
$\bm m = \bm\al^*$, $\bm\al\in\Phi(G)$, which have masses~\cite{bais}
\beq
\left(M_m\right)_{\pm\al} = \frac{4\pi}{e} \mod{\bm f \cdot \bm\al^*}.
\eeq
Comparing this to the gauge bosons (\ref{mgauge}) gives 
\bea
\label{mrel}
\left(M_W\right)_{\pm\al^*}=\left(M_m\right)_{\pm\al},\hspace{2em}
\left(M_m\right)_{\pm\al^*}=\left(M_W\right)_{\pm\al},\\
\left(\bm E_W\right)_{\pm\al^*}=\left(\bm B_m\right)_{\pm\al},\hspace{2em}
\left(\bm B_m\right)_{\pm\al^*}=\left(\bm E_W\right)_{\pm\al}
\eea
providing the gauge coupling $g$ of $G^*$ satisfies $eg=4\pi N$.
Then the gauge bosons of $G^*\rightarrow H^*$ have the same charges and
masses as the above monopoles in $G\rightarrow H$. Implications include:\\
(a) The monopoles with $\bm m=\bm\al^*$ form charge multiplets identical with
the gauge bosons of $G^*\rightarrow H^*$.\\
(b) This gives good evidence for the Goddard, Nuyts and Olive conjecture
that $G\rightarrow H$ monopoles form charge multiplets and interact under
the dual group $H^*$.

Note these monopoles have no 
intrinsic angular momentum, while gauge bosons are spin one. Therefore
an exact duality between monopoles and gauge bosons is not expected 
(unless supersymmetry is included). Here we interpret (a) and (b)
as evidence for identifying the multiplet structures. 
Furthermore this identification holds for non-BPS monopoles
because their multiplet structures depend only on their charges~\cite{com1}.

To proceed, we make a useful definition. Writing
\beq
\Phi(G)=\Phi(H)+\Phi(\calM)
\eeq
divides the gauge bosons into massless and massive sets. Then the massive
monopoles have $\bm m=\bm\ga^*$, $\bm\ga\in\Phi(\calM)$; while the
massless dual roots are associated with gauge transforming these monopoles.
By points (a) and (b) above these massive monopoles form a 
representation of $H^*$ identical to the massive gauge bosons of 
$G^*\rightarrow H^*$.

This observation 
relates to the monopoles in tables~\ref{tab1} to \ref{tab2} because
each $n=1$ monopole has $\bm m=\bm\ga^*$ with $\bm\ga\in\Phi(\calM)$.
That $\bm m=\bm\ga^*$ can be seen directly by calculating the roots, although 
one could also note the $n>1$ monopoles are composites of the $n=1$ monopoles,
which are therefore the fundamental monopoles associated with simple 
roots~\cite{wein}.

Then Bais' generalization of the Montenon-Olive conjecture can be applied
to the monopole-elementary
particle correspondences in tables~\ref{tab1} to \ref{tab2}. For this 
application, we determine the representations of the $n=1$ monopoles 
and compare these with the elementary particles.

First, note the $\su(n)$ algebras are self-dual~\cite{com2}
\beq
\su(n)^*=\su(n).
\eeq
Then the representation of
the massive dual gauge bosons under the algebra $\calH^*$ is identical
to that of the massive gauge bosons under $\calH$.

Second, note the gauge-elementary particle interactions (vertices)
depend on the $\su(3)_\rmC\oplus\su(2)_\rmI\oplus\u(1)_\rmY$ representation,
with the elementary particles
taking fundamental representations of this algebra.

Then the representations of the monopoles are
consistent with the elementary particles if the
massive gauge bosons in (\ref{unif}) to (\ref{su2}) take fundamental 
representations:

\noindent (i) $\su(5)\rightarrow\su(3)\oplus\su(2)\oplus\u(1)$:
Labeling the twelve roots of $\Phi(\calM)$
by $\pm\bm\ga(i,j)$ (i=1,2,3; j=1,2) gives in component form
\beq
(E^{+\ga})_{ab}=\de_{ai}\de_{b,j+3},\hspace{2em}
(E^{-\ga})_{ab} = \de_{a,j+3} \de_{bi}.
\eeq
Defining $(e^\ga)_{ab}=\de_{ai}\de_{bj}$, we then have for any algebra 
element $X\in\su(3)\oplus\su(2)\oplus\u(1)$:
\beq
\label{eq}
\ad(X)W_\mu = \ad(X) \sum_{\pm\ga} W_\mu^\ga E^\ga
= X\,\sum_{+\ga} W_\mu^\ga e^\ga,
\eeq
where $\su(3)$ acts from the left and $\su(2)$ the right. Clearly, (\ref{eq})
forms a fundamental representation.

Therefore the $n=1$ monopole in table~\ref{tab1} takes a fundamental
representation of $\su(3)\oplus\su(2)\oplus\u(1)$, in agreement with the
corresponding $(u,d)_L$ particle multiplet.

\noindent (ii) $\su(4)\rightarrow\su(3)\oplus\u(1)$:
Labeling the six roots of $\Phi(\calM)$ by $\pm\bm\ga(i)$ (i=1,2,3)
gives in component form
\beq
(E^{+\ga})_{ab}=\de_{ai}\de_{b,4},\hspace{2em}
(E^{-\ga})_{ab} = \de_{a,4} \de_{bi}.
\eeq
Defining $(e^\ga)_a=\de_{ia}$, we then have
for $X\in\su(3)\oplus\u(1)$ equation~(\ref{eq}) holding with
$\su(3)$ acting from the left. Again this forms a fundamental representation.

Thus the $n=1$ monopole in table~\ref{tab4} takes a fundamental
representation of $\su(3)\oplus\u(1)$ in agreement with the corresponding
$\bar d_L$ elementary particle.

\noindent (iii) $\su(3)\rightarrow\su(2)\oplus\u(1)$:
Labeling the four roots of $\Phi(\calM)$ by $\pm\bm\ga(j)$ (j=1,2) gives
\beq
(E^{+\ga})_{ab}=\de_{aj}\de_{b,3},\hspace{2em}
(E^{-\ga})_{ab} = \de_{a,3} \de_{bj}.
\eeq
Defining $(e^\ga)_a=\de_{ja}$, we then have
for $X\in\su(2)\oplus\u(1)$ equation~ (\ref{eq}) holding with
$\su(2)$ acting from the left, which again forms a fundamental representation.

Thus the $n=1$ monopole in table~\ref{tab3} takes a fundamental
representation of $\su(2)\oplus\u(1)$ in agreement with 
the $(\bar\nu, \bar e)_R$ elementary particle multiplet.

\noindent (iv) $\SU(2)\rightarrow\U(1)$: This time there are only
two roots with matrices
\beq
E^+ = \left( \begin{array}{cc} 0&1\\ 0&0 \end{array} \right),\hspace{2em}
E^- = \left( \begin{array}{cc} 0&0\\ 1&0 \end{array} \right).
\eeq
Then for $h(\theta)=\diag\left(e^{i\theta},e^{-i\theta}\right)\in\U(1)$
we have
\beq
\Ad(h)W_\mu=
\left(
\begin{array}{cc}0&e^{i\theta}W^+_\mu\\(e^{i\theta}W^+_\mu)^\dagger&0
\end{array}\right),
\eeq
which forms a fundamental representation of $\U(1)$.
Thus the $n=1$ monopole in table~\ref{tab2} takes a fundamental
representation of $\u(1)$, in agreement with $\bar e_L$.

In conclusion, we described some simpler versions of the
monopole-elementary particle correspondence that underlies the dual
standard model. These correspondences are summarized in tables~\ref{tab1} 
to \ref{tab2}. Then we discussed the relation of these correspondences
to Bais' generalization of 
the Montenon-Olive conjecture ---
finding in particular that the representation of each $n=1$ monopole is
consistent with the corresponding elementary particle multiplet.

Some additional comments are:

\noindent (a) The arguments given in this letter apply only to the $n=1$ 
monopoles in tables~\ref{tab1} to \ref{tab2}.
Note they do not apply to the $\bar u_R$ correspondence.
An interesting coincidence is that all SU(5) monopoles in table~\ref{tab1}
are spherically symmetric apart from the one 
associated with $\bar u_R$.

\noindent (b) As mentioned before, classical monopoles have no intrinsic
angular momentum and so we apply Bais' results only to the representations 
of the dual gauge bosons and monopoles (since we are not considering
supersymmetry).

Similarly, the elementary particles are spin-half; therefore our discussion
is again limited only to the representations.
However, to construct a dual standard model 
it is necessary to obtain classical solutions with one-half intrinsic
angular momentum. This can be achieved by considering dyons, which have
an angular momentum 
\beq
\bm J = \int d^3r\,\,\bm r \wedge (\bm E \wedge \bm B).
\eeq
Vachaspati~\cite{vach95} suggested forming monopole-scalar boson 
composites similar to the $J=\half$ states discussed by Hasenfratz and 
't~Hooft, and Jackiw and Rebbi~\cite{hasen}.

\noindent (c) We comment that each monopole spectrum in tables
\ref{tab4} to \ref{tab2} isolates a specific feature of the SU(5)
monopole-elementary particle correspondence in table~\ref{tab1}.

The SU(3) monopole spectrum isolates the electroweak symmetry. This may be 
of use for studying electroweak symmetry breaking in the dual standard 
model.

Similarly the SU(4) monopole spectrum isolates the 
strong and hypercharge symmetries. This may be of use for studying 
confinement in the dual standard model. For instance, confinement
in the dual standard model can be described by breaking dual 
colour~\cite{vach95}; such a feature also occurs in the SU(4) model.

\noindent (d) Finally we mention that the application of non-Abelian duality
to the dual standard model is but one example of applying traditional
properties of gauge theories to the particle-monopole correspondence in 
table~\ref{tab1}.
For instance, Nambu's description of confinement also applies.
Furthermore, Skyrme described nucleons as classical configurations of the pion 
field --- 
the dual standard model is the same concept applied to gauge theories.

\noindent {\em Endnote: the present discussion is intended to 
supersede the previous one by this author in ref.~\cite{m}, which discusses
how \/$\SU(5)$ monopoles transform under an electric 
$\SU(3)\x\SU(2)\x\U(1)/\bbZ_6$ gauge symmetry. To properly examine 
a duality between $\SU(5)$ monopoles and the elementary particles, one should 
compare the magnetic gauge freedom of the monopoles with the electric gauge 
freedom of the elementary particles.}

\footnotetext[1]{email: n$\_$lepora@hotmail.com}

\begin{thebibliography}{99}

\bibitem{vach95}
T.~Vachaspati,
Phys.\ Rev.\ Lett.\ {\bf 76} (1996) 188
[hep-ph/9509271];
H.~Liu and T.~Vachaspati,
Phys.\ Rev.\ {\bf D56} (1997) 1300
[hep-th/9604138];
A.~S.~Goldhaber,
Phys.\ Rept.\  {\bf 315} (1999) 83
[hep-th/9905208];
T. Vachaspati, Phys.\  Lett.\  {\bf B427} (1998) 323 
[hep-th/9709149].

\bibitem{meunif}
N.~F.~Lepora,
JHEP {\bf 0002} (2000) 036
[hep-ph/9910493].

\bibitem{bais}
F.~A.~Bais,
Phys.\ Rev.\ D {\bf 18} (1978) 1206.

\bibitem{mont}
C.~Montonen and D.~I.~Olive,
Phys.\ Lett.\ B {\bf 72} (1977) 117.

\bibitem{gno}
P.~Goddard, J.~Nuyts and D.~I.~Olive,
Nucl.\ Phys.\ {\bf B125} (1977) 1.

\bibitem{gard84}
C.~Gardner and J.~Harvey,
Phys.\ Rev.\ Lett.\ {\bf 52} (1984) 879.

\bibitem{engle}
F.~Englert and P.~Windey,
Phys.\ Rev.\ D {\bf 14} (1976) 2728.

\bibitem{com1} Note the mass relations~(\ref{mrel}) only hold when BPS. 
However there are remnants of this relation for non-BPS
monopoles. For instance, two gauge bosons defined by $\bm\al$ and 
$\bm\al+\bm\de$ with $\bm\de\in\Phi(H)$ are of equal mass; while two
monopoles defined by $\bm\al^*$ and $\bm\al^*+\bm\de^*$
have equal mass. This behaviour is consistent with $H$ and $H^*$ gauge 
invariance.

\bibitem{wein}
E.~J.~Weinberg,
Nucl.\ Phys.\  {\bf B167} (1980) 500;
E.~J.~Weinberg,
Nucl.\ Phys.\  {\bf B203} (1982) 445.

\bibitem{com2} This is because all roots of $\su(n)$ have the same
length. The dual group (\ref{H*}) is defined by swapping the long and
short roots.

\bibitem{hasen}
P.~Hasenfratz and G.~'t Hooft,
Phys.\ Rev.\ Lett.\  {\bf 36} (1976) 1119;
R.~Jackiw and C.~Rebbi,
Phys.\ Rev.\ Lett.\  {\bf 36} (1976) 1116.

\bibitem{m}
N.~F.~Lepora,
JHEP {\bf 0002} (2000) 037.

\end{thebibliography}
\end{document}